\magnification = \magstep1
\baselineskip=14pt
\parskip= \smallskipamount
\newcount\ftnumber
\def\ft#1{\global\advance\ftnumber by 1
          {\baselineskip 11pt    
           \footnote{$^{\the\ftnumber}$}{#1}}}
\def\abst#1{\medskip{\baselineskip=12pt
{\narrower\narrower\parindent = 0pt #1 \par}}} 

\centerline{{\bf Nonlocality and Bohr's reply to EPR}} \bigskip
\centerline{N. David Mermin} \centerline{Laboratory of Atomic and Solid
State Physics} \centerline{Cornell University, Ithaca, NY 14853-2501}
\medskip

\abst{Henry Stapp's commentary (quant-ph/9711060) does not capture the
point I was trying to make in my essay (quant-ph/9711052) on how a
subtle flaw in his ``proof of quantum nonlocality'' clearly
illustrates a central issue in Bohr's reply to EPR. I therefore wish
to emphasize what I do and do not say in that essay and even,
with some trepidation, what Bohr did and did not say in his reply to
EPR.}  \bigskip

In examining Henry Stapp's reading (quant-ph/9711060) of my use
(quant-ph/9711 052)\ft{Note that on the second line of page 7 of my
manuscript, $R1-$ should (clearly) be $R1+$.} of his beautiful but
flawed nonlocality argument\ft{Henry P.~Stapp, ``Nonlocal character of
quantum theory,'' Am.~J.~Phys.~{\bf 65}, 300-304 (1997).} to elucidate
Bohr's reply to EPR, I was strongly reminded of the importance of
utmost caution in all questions of terminology and
dialectics.\ft{Neils Bohr, in {\it Albert Einstein,
Philosopher-Scientist}, P.~Schilpp ed., Open Court, La Salle,
Illinois, 1949, p.~237.} Several cautionary remarks are in order.

1.  Stapp refers to ``the nonlocal influence deduced by Bohr''
[Abstract]\ft{Bracketed references are all to Stapp's quant-ph/9711060.}.  He
talks about ``the faster-than-light influence that [Bohr] claims
exists'' [p.~5] and the ``action-at-a-distance whose existence [Bohr]
claimed'' [p.~7].   Nowhere, however, do the terms ``nonlocal'' or
``faster-than-light'' or ``action-at-a-distance'' appear in Bohr's reply
to EPR, nor, so far as I know, can they be found in any of his
subsequent writings.  Such terms hardly seem suited to characterize the
influence that Bohr does identify: an influence on the possibility of
making valid predictions.  The influence established by Stapp's argument
has a similarly insubstantial character: it is an influence on the
possibility of making counterfactual statements that meet his own
criterion for their valid use.  

2. The statement Stapp says I grant [p.~1], under the condition that
experiment L2 is performed on the left, differs significantly from my
own translation of his modal-logical symbolism into ordinary
language.  What I would accept is ``(S): If performing R2 does give +,
then performing R1 instead would give --.'' Stapp's version uses
``were to'' instead of ``does'', and ``must'' instead of ``would'',
thereby blurring the crucial distinction between what actually happened
(performing R2 and getting the result +) and what might have happened
but did not (performing R1).  This distinction must be kept sharp
because what I do grant is that that the counterfactual part of the
statement (which must, of course, be clearly identifiable) meets
Stapp's criterion under the stated condition. 

3.  When a statement about what would have happened in an unperformed
experiment fails to meet Stapp's criterion it is not ``false'' [p.~1]
--- it is meaningless, be it ever so ``well-formed'' [p.~6].  When I
insist that such a counterfactual has no meaning in the context of
Stapp's argument I am not ``capriciously'' adopted a ``rule'' or
engaging in an ``evasion of logic'' or ``arbitrarily limiting the scope
of logical reasoning'' [p.~7].  I am merely noting that the
statement fails to meet Stapp's own rule for when a counterfactual can
make sense.  

4.  The question I raise is not whether ``LOC2 fails'' [p.~1].  My
point is that Stapp uses his locality principle LOC2 in a case where
the conditions for its application have not been met.  The essential
ambiguity responsible for the gap in Stapp's argument lies not in LOC2
but in the statement (S) to which Stapp wishes to apply LOC2.  The
counterfactual part of (S), which apparently refers only to past events
(which is essential for the applicability of LOC2), actually makes
implicit reference to future events, through Stapp's criterion for the
validity of a counterfactual statement.  This implicit future
reference does not make LOC2 {\it fail\/}; it renders LOC2 {\it
inapplicable\/}.

5.  I do indeed understand, and Stapp's original argument makes it
quite clear, that his use of LOC2 is designed precisely to circumvent
the fact that the direct proof of (S) requires the experiment L2 to be
performed [p.~2].  My point is not that he overlooks the importance of
L2 being performed. It is that his stratagem for getting around its
non-performance with the help of LOC2 fails, because --- and this is
what he does overlook --- LOC2 is then inapplicable, for the reason
noted in 4 above.  

6.  If ``truth of'' were replaced by ``possibility of making sense of''
then I would agree that ``the issue $\ldots$ is whether a nontrivial dependence
of the truth of (S) upon a future free choice means that there is some
sort of backwards-in-time influence'' [p.~2].  In addressing that
issue I would maintain that because the influence is only on the
possibility of satisfying Stapp's criterion for the valid use of
counterfactuals, it is inappropriately characterized as
``backwards-in-time'', or ``faster-than-light'', or ``nonlocal''. 

7.  An appealing feature of Stapp's original argument was that it
eschewed all incautious talk of such problematic notions as ``Nature's
choices'', ``Nature's earlier selections'', or even ``properties''.
Such terminology reappears in his commentary [pp.~2,3].  If these
notions are essential to his argument, this raises other issues and,
it seems to me, makes that argument less interesting.  If (as I
suspect) he has reintroduced them only to provide a more intuitive
feel for the kind of ``nonlocality'' he claims to have derived, then
they have no relevance to my point, which concerns the internal
coherence of his original argument.

8. Stapp correctly notes that I question ``whether the fact that
statement (S) refers explicitly only to possible measurements and
possible results confined to the right-hand region $R$ really
justifies'' Stapp's crucial step [p.~5].  But I do so not from an
excessive preoccupation with his original proof of (S), but because
the absence of an {\it explicit\/} reference to the left does not
preclude the presence of an {\it implicit\/} reference, as
noted in 4 above.  To give (S) meaning Stapp's criterion requires
the actual performance on the left of an experiment whose result makes
it possible to deduce from the result of the actual experiment on the
right, what the result would have been for the experiment on the right
that was not actually performed.

9.  The reason I get ``involved in questions of definitions and
meanings'' [p.~5] is that Stapp himself gives a clear criterion for
when a counterfactual can have meaning in a physical argument, and
because the meaning of a counterfactual statement is central in the
step in his argument that I examine.  As I read Bohr, he too was broadly
interested in questions of definition and meaning. Their appearance in
his reply to EPR is similar to their appearance in my discussion of
Stapp's nonlocality argument.  As I note toward the end of my essay,
Stapp's crucial counterfactual and EPR's crucial
prediction-with-certainty play interestingly similar roles, except for
a reversal of the time order of the relevant events.  As a result of
this exchange of predictions and counterfactuals, Stapp's argument
provides a more transparent illustration of what I (but not Stapp)
believe to be the nature of Bohr's reply to EPR. 

10. Lucien Hardy and John Bell before him fatally undermine the
position of EPR.  But they make surprisingly little difference to the
internal coherence of Bohr's side of the exchange.  Whether Bohr knew
in his bones that there were no elements of reality or knew it through
backwards-in-time messages from Bell and Hardy, has no bearing on what
he criticizes in EPR's reasoning.  I agree that ``the Hardy-based
analysis fortifies Bohr's position'' [p.~7, Abstract], but only
because it makes one take seriously the urgent need to find a flaw in
the apparently cogent reasoning of EPR.  This is why the flaw that
invalidates Stapp's ingenious attempt to derive nonlocality --- the
alternative EPR's reasoning leads to if elements of reality
are excluded (as they themselves remark) --- can be so similar, even
after Bell and Hardy, to the one Bohr identified over sixty years ago.

\bigskip

{\sl Acknowledgment.} These remarks relate to work supported by the
National Science Foundation under Grant No.~PHY9722065.

\bye